\documentclass[a4paper]{jpconf}
\usepackage{graphicx}
\begin{document}
\pagestyle{plain}
\title{CEM03 and LAQGSM03---new modeling tools 
for nuclear applications}

\author{S G Mashnik$^1$, A J Sierk$^1$, K K Gudima$^2$ and M I Baznat$^2$}

\address{$^1$  Los Alamos National Laboratory, Los Alamos, NM 87545, USA\\
 $^2$ Institute of Applied Physics, Academy of Science of Moldova,
Chi\c{s}in\u{a}u, Moldova}

\ead{mashnik@lanl.gov}

\begin{abstract}
An improved version of the Cascade-Exciton Model (CEM) of nuclear 
reactions realized in the code CEM2k and the Los Alamos version of 
the Quark-Gluon String Model (LAQGSM)
have been developed recently at LANL
to describe reactions induced by particles and nuclei 
for a number of applications. 
Our CEM2k and LAQGSM merged with the GEM2 evaporation/fission
code by Furihata have predictive 
powers comparable to other modern codes and describe many 
reactions better than other codes; therefore both our codes can be used
as reliable event generators in transport codes for applications. 
During the last year, we have made a significant improvements to the 
intranuclear cascade parts of CEM2k and LAQGSM, and have extended LAQGSM to
describe photonuclear reactions at energies to 10 GeV and higher. We have
produced in this way improved versions of our codes, CEM03.01 and 
LAQGSM03.01. For special studies, we have also merged our two
codes with the 
GEMINI code by Charity and with the 
SMM
code of Botvina.
We present a brief description of our codes and show illustrative 
results obtained with CEM03.01 and LAQGSM03.01 for different reactions
compared with predictions by other models, as well as
examples of using our codes as modeling tools for nuclear
applications.
\end{abstract}

\section{Introduction}
Following an increased interest in 
nuclear data 
for such projects as the Accelerator Transmutation of nuclear 
Wastes (ATW), 
Accelerator Production of Tritium (APT), 
Spallation Neutron Source (SNS), 
Rare Isotope Accelerator (RIA), 
Proton Radiography (PRAD) as a radiographic probe for the Advanced 
Hydro-test Facility and others, for several years the US Department 
of Energy has supported our work on the development of 
improved versions of the Cascade-Exciton Model (CEM)
of nuclear reactions \cite{CEM2k,JNRS05,CEM03.01}
and the Los Alamos version of the Quark-Gluon String Model (LAQGSM) 
\cite{LAQGSM}
to describe reactions induced by particles and
nuclei at energies up to about 1 TeV/nucleon.
To describe fission and production of light fragments heavier
than $^4$He, we have
merged both our codes with several evaporation/fission/fragmentation models,
including 
the Generalized Evaporation/fission Model code
GEM2 by Furihata \cite{GEM2}. Our codes
perform as well as and often better than other current 
models in describing a large variety of spallation,
fission, and fragmentation reactions, therefore they are used
as event-generators in several transport codes.
The status of our codes as of the middle of 2004 together with
illustrative results and comparisons with other models can be found
in \cite{ND2004,astro,TRAMU} and references therein. Here, we present 
several improvements developed during the last year leading
to the new versions of our codes, CEM03.01 and LAQGSM03.01
\cite{CEM03.01,ResNote05}.

\section{Basic assumptions of CEM and LAQGSM}
The Cascade-Exciton Model (CEM) of nuclear reactions 
was proposed initially at the
Laboratory of Theoretical Physics, JINR, Dubna \cite{CEM}
to describe intermediate-energy spallation reactions induced by
nucleons and pions. 
It is based on the Dubna IntraNuclear Cascade (INC) \cite{Book}
and the Modified Exciton Model (MEM) \cite{MEM,MODEX}.
It was extended later to consider photonuclear reactions 
and to describe fission cross sections
using different options for nuclear masses, fission barriers, 
and level densities, 
and its 1995 version CEM95 was released to the public via NEA/OECD, Paris 
and RSICC, Oak Ridge (see corresponding references in \cite{CEM03.01}).

LAQGSM is an extension by Gudima, Mashnik and Sierk
of the Quark Gluon-String Model (QGSM) 
by Amelin, Gudima and Toneev \cite{QGSM}
done at Los Alamos  \cite{LAQGSM} and is intended to describe
both particle- and nucleus-induced reactions at energies up to
about 1 TeV/nucleon. LAQGSM is based on
a time-dependent version of the
intranuclear-cascade model (different from the one used in CEM)
developed at JINR, Dubna, often referred to in the literature simply
as the Dubna intranuclear Cascade Model (DCM) (see \cite{DCM}
and references therein).
LAQGSM \cite{LAQGSM,ResNote05} differs from QGSM \cite{QGSM} by 
using an extended and significantly improved version of the
INC model \cite{ND2004,ResNote05}, by
replacing the preequilibrium and
evaporation parts of QGSM described according to the standard CEM 
\cite{CEM} with the new physics from CEM03.01 
\cite{CEM03.01} and also has a number of improvements 
and refinements in the Fermi break-up and coalescence models 
in comparison with QGSM \cite{QGSM}.

CEM and LAQGSM assume
that the reactions occur generally in three stages. 
The first stage is the INC,
in which primary particles can be re-scattered and produce secondary
particles several times prior to absorption by, or escape from the nucleus.
When the cascade stage of a reaction 
is completed, CEM and LAQGSM use the coalescence model
as described in \cite{DCM}
to ``create" high-energy d, t, $^3$He, and $^4$He by
final-state interactions among emitted cascade nucleons.
The emission of the cascade particles determines the particle-hole 
configuration, Z, A, and the excitation energy that is
the starting point for the second, preequilibrium stage of the
reaction.  
The subsequent relaxation of the nuclear excitation is
treated in terms of an improved version of the modified exciton 
model of preequilibrium decay 
followed by the equilibrium evaporation/fission stage of the reaction.
Generally, all four components may contribute to experimentally measured 
particle spectra and other distributions. 
But if the residual nuclei after the INC have atomic numbers 
with  $A \le 12$,  CEM03.01 and LAQGSM03.01 use the Fermi 
break-up model \cite{Fermi:50} to calculate their further disintegration 
instead of using the preequilibrium and evaporation models. 
Fermi break-up is much faster to calculate and gives results very similar 
to the continuation of the more detailed models to
much lighter nuclei.

\section{Recent developments in CEM03.01 and LAQGSM03.01 and
illustrative results} 

First, during 2004 we incorporated into LAQGSM 
the improved approximations for the total elastic and inelastic
cross sections of hadron-hadron and photo-hadron elementary
interactions developed previously for the code CEM97 
\cite{CEM97} and CEM2k \cite{CEM2k}
(see details in \cite{CEM97}).

Second,  the double differential distributions of
secondary particles from elementary
$NN$ and $\gamma N$ interactions were simulated by CEM2k
(and all its precursors, as well as by LAQGSM and its precursors
at energies below 4.5 GeV/A) 
still using the old Dubna INC \cite{Book} approximations 
that were obtained by Gudima {\it et al.} \cite{JINR68}
36 years ago, using the measurements available at that time.
For instance, in the case of two-body reactions, 
the cosine of the angle of emission of secondary particles 
in the c.m. system is calculated by the Dubna INC
as a function of a random number $\xi$, distributed uniformly in the 
interval [0,1] as
\begin{equation}
\cos \theta = 2 \xi ^{1 / 2} \left[ \sum_{n=0}^{N} a_n \xi^n
+ (1- \sum_{n=0}^N a_n ) \xi^{N+1} \right] -1 \mbox{ ,}
\end{equation}
where $N = M = 3$,
\begin{equation}
a_n = \sum_{k=0}^M a_{nk} T_{i}^k \mbox{ .}
\end{equation}
The coefficients $a_{nk}$ were fitted to the then available 
experimental data at a number of
incident kinetic energies $T_i$ , then interpolated and extrapolated
to other energies (see details in \cite{Book,JINR68} and references
therein).
The distribution of secondary particles over the azimuthal angle
$\varphi$ is assumed isotropic. For elementary interactions 
with more than two particles in the final state, 
the Dubna INC uses the statistical model to simulate the angles 
and energies of products (see details in \cite{Book}).

For the improved versions of our codes
referred to as CEM03 and LAQGSM03, respectively,
we use currently available experimental data and recently published
systematics proposed by other authors to develop new
approximations for angular and energy distributions of
particles produced in nucleon-nucleon and photon-proton interactions.
So, for $pp$, $np$, and $nn$ interactions at energies up to
2 GeV, we did not have to develop our own approximations
analogous to the ones described by Eqs.\ (1) and (2),
since reliable systematics have been developed recently
by Cugnon {\it et al.} for the Liege INC \cite{INCL}, then 
improved further by Duarte for the BRIC code \cite{BRIC1.4};
we simply incorporate into CEM03 and LAQGSM03 the 
systematics by Duarte \cite{BRIC1.4}. Similarly, for
$\gamma N$ interactions, 
we take advantage of the event generators for $\gamma p$ and
$\gamma n$ reactions from the Moscow INC \cite{Iljinov97}
kindly sent us by Dr.\ Igor Pshenichnov. 
In our codes, we use part of a large data file with smooth 
approximations through presently available experimental data from 
the Moscow INC \cite{Iljinov97} and have developed 
a simple and fast algorithm to simulate unambiguously
$d \sigma / d \Omega$ and to
choose the corresponding value of $\Theta$ for any  $E_\gamma$,
using a single random number $\xi$ uniformly
distributed in the interval [0,1] \cite{JNRS05}. 
For other elementary interactions, we fit new sets of parameters 
$a_n$ from Eq.\ (1) at different $T_i$ for which we found data, then
we approximated the energy dependences of the parameters 
$a_{nk}$ in Eq.\ (2) using the fitted values of $a_n$.

Examples of angular distributions of secondary particles from
$np$ and $\gamma p$ reactions at several energies are shown
in Figs.\ 1 and 2 of Ref.\ \cite{ND2004}. The new approximations
from CEM03 and LAQGSM03 reproduce the experimental data
much better than the old Dubna INC used in our previous code 
versions (and in several other codes developed from the 
Dubna INC) and allow us to describe better particle spectra
from different reactions on nuclei (see, {\it e.g.}, Fig.\ 3 in \cite{ND2004}).

Third, we have improved 
the description of complex-particle spectra in CEM03 and LAQGSM03.
This was done by refining the coalescence model used in our codes, 
by developing a better approach to estimate the probability of 
complex-particle emission at the preequilibrium stage of a reaction,
and by incorporating into our codes the known systematics for
angular distributions of complex particles developed
by Kalbach (see details in \cite{CEM03.01}).

Fig.\ 1 shows examples of proton, deuteron, and triton spectra
from 542 MeV p + Cu and Bi calculated by the recently improved version
of CEM compared with experimental data \cite{Franz90}. We
see that although the CEM03.01 results do not coincide exactly
with these experimental data, 
the agreement is comparable to that
provided by other modern models like FLUKA (see Fig.\ 3 in \cite{FLUKA})
and the latest version of the Liege INC by Cugnon {\it et al.} 
(see Fig.\ 3 in \cite{INCL5}), and is significantly better than that 
obtained a decade ago with the initial version of CEM \cite{CEM}
(see  Figs.\ 6, 7, 9 and 10 in \cite{NP94}).

We note that CEM03.01 also describes reasonably well complex-particle 
(and nucleon) spectra from reactions at energies 
below 100 MeV, where more sophisticated microscopic codes 
like GNASH \cite{GNASH} or TALYS \cite{TALYS} are usually
used to produce data libraries for applications.
For example, from Fig.\ 6 of Ref.\ \cite{CEM03.01} one can see that
CEM03.01 describes the nucleon and complex-particles spectra
from 62.9 MeV p + $^{208}$Pb measured recently \cite{Guertin05}
in Louvain-la-Neuve for the HINDAS project quite well and agrees with
the data about as well as results by TALYS \cite{TALYS}, 
GNASH \cite{GNASH}, FLUKA \cite{FLUKA}, or 
INCL4 \cite{INCL} published in Figs.\ 20 and 21
of Ref.\ \cite{Guertin05}.

\begin{figure}[h]
\begin{minipage}{19.5pc}

\includegraphics[width=19.5pc]{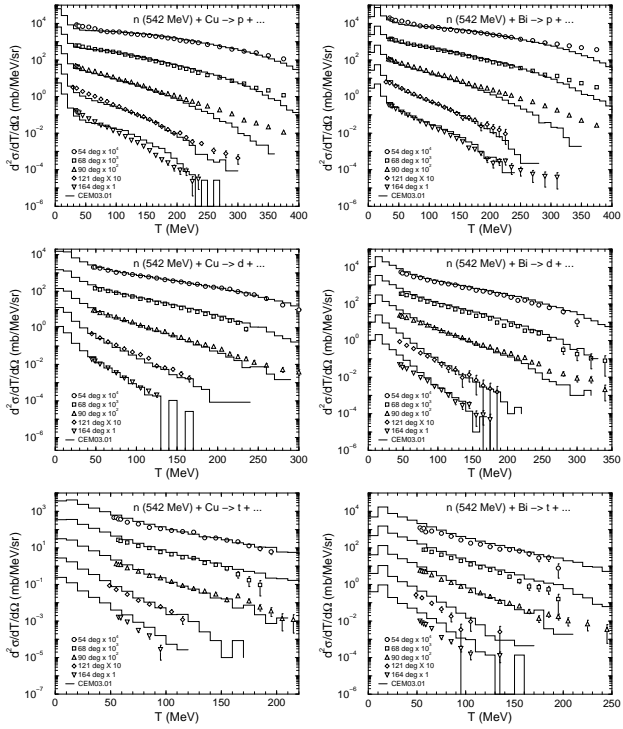}
\caption{\label{label}
Measured \cite{Franz90} and calculated proton, deuteron,
and triton spectra from 542 MeV p + Cu and Bi, respectively.}

\end{minipage}
\hspace{1.0pc}%
\begin{minipage}[h]{17.5pc}

Next we extended CEM03 to calculate reactions induced
by both monochromatic and bremsstrahlung photons, as described
in detail in \cite{JNRS05}. As one can see from the example
shown in Fig.\ 2 and from many other results published in
\cite{JNRS05,CEM03.01}, CEM03.01 now describes reasonably
well photonuclear reactions at incident energies from about
30 MeV to $\sim 1.5$ GeV. It can be used as well for estimation
of photonuclear reactions for applications at higher
energies, up to $\sim 5$ GeV (see details in \cite{JNRS05,CEM03.01}).

Finally, 
after making the improvements to the INC and preequilibrium parts 
CEM (and LAQGSM) as described above, the mean values of the 
mass and charge numbers, $A$ and $Z$ of the excited compound nuclei
produced after the preequilibrium stage of nuclear reactions and their
mean excitation energy $E^*$ have changed slightly, which
affects the probability of heavy compound nuclei
(especially preactinides) to fission. This means that
the procedure of fitting the ratio of the level-density
parameters of fissioning nuclei at the saddle point, $a_f$, and
for evaporation from compound 
nuclei, $a_n$, 
\end{minipage}
\end{figure}

\begin{figure}[h]
\begin{minipage}{23pc}
\hspace{2.5pc}
\includegraphics[width=16.5pc]{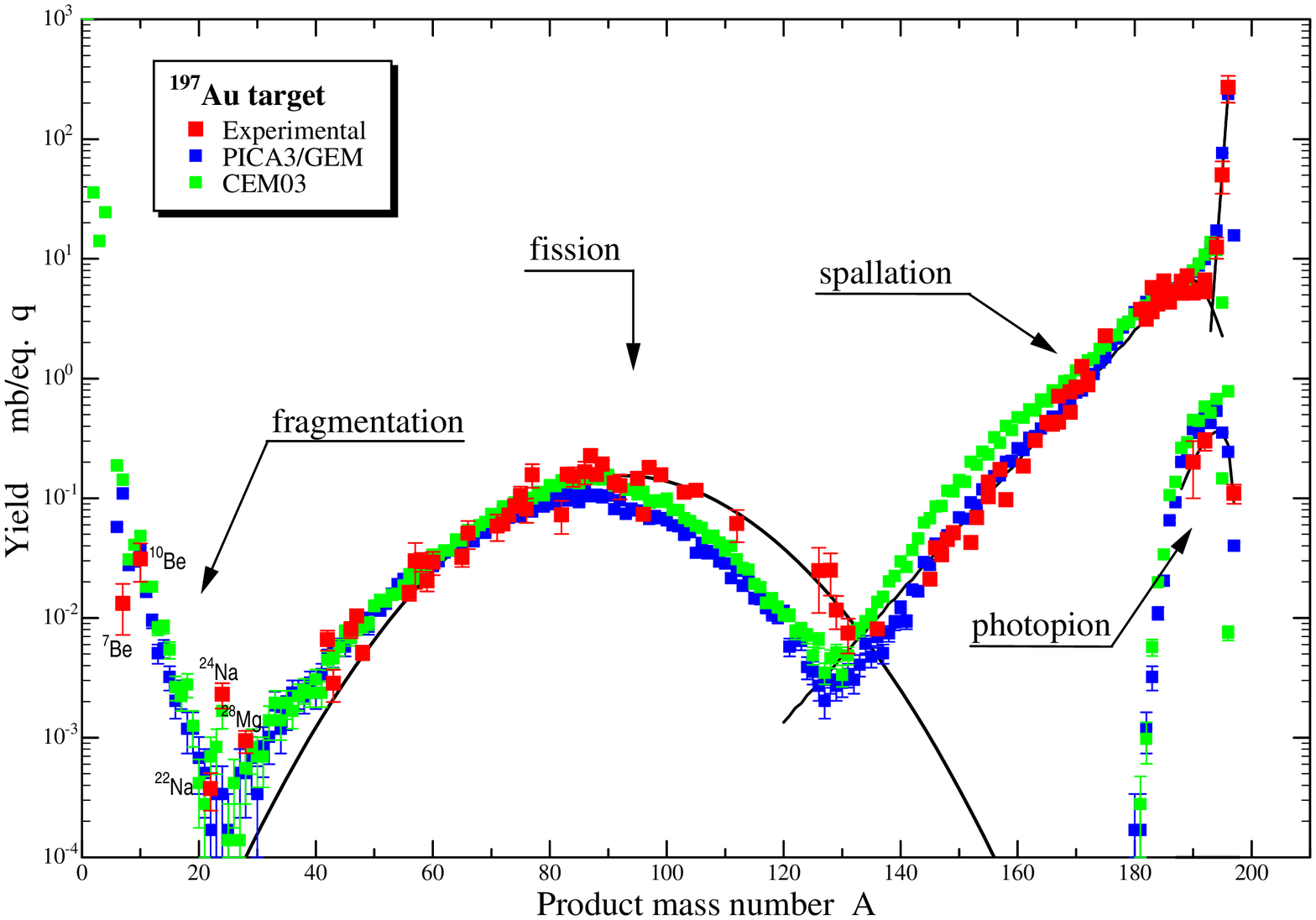}
\caption{\label{label}
CEM03.01 results (green symbols)
for the isotopic yields of products from $E_0 = 1$ GeV
bremsstrahlung-induced
reactions on $^{197}$Au compared
with experimental data (red symbols) \cite{Sakamoto03} 
and calculations by PICA3/GEM (blue symbols);
the PICA3/GEM results are from several publications and
are presented in Fig.\ 18 of \cite{Sakamoto03}. 
The 
black curves represent approximations based on 
experimental data 
\cite{Sakamoto03}
(see more details in \cite{JNRS05}).
}
\end{minipage} 
\hspace{0.5pc}
\begin{minipage}{13.8pc}

\vspace*{-4mm}
\hspace{-5mm}
\includegraphics[width=15.pc]{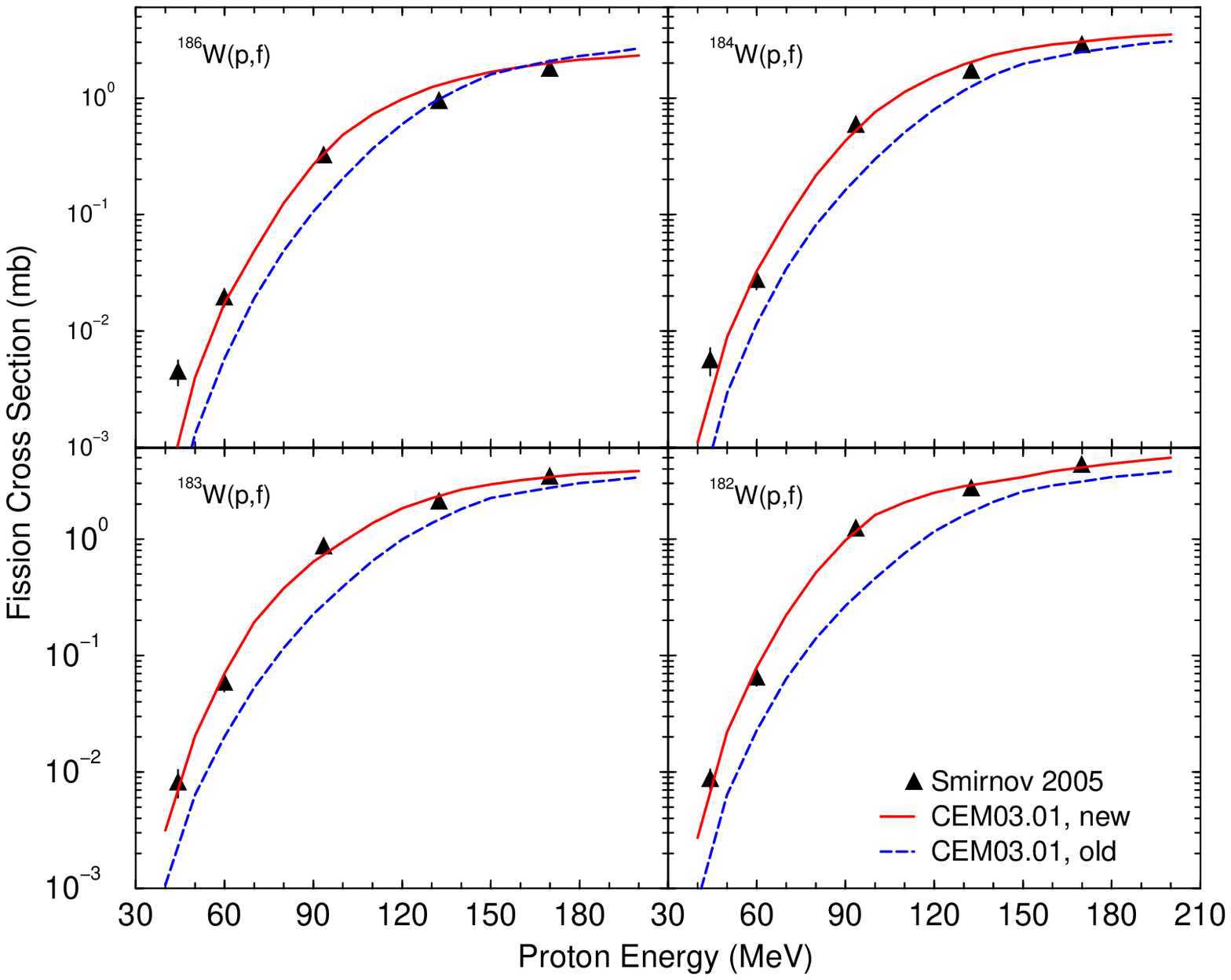}
\caption{\label{label}
Experimental \cite{Eismont05}
proton-induced fission cross sections of $^{186}$W, $^{184}$W, $^{183}$W,
and $^{182}$W compared with improved (red solid lines) and old
(blue dashed lines, from \cite{Eismont05}) CEM03.01 calculations.
}
\end{minipage}\hspace{1.5pc}%
\end{figure}
{\noindent
{\it i.e.}, $a_f/a_n$ which we performed in \cite{fitaf} to provide the 
best description by CEM2k and LAQGSM
of fission cross sections is no longer correct.}
We redo for CEM03.01 and  LAQGSM03.01 the determination of the 
level-density ratio done in  \cite{fitaf} for CEM2k and LAQGSM, 
ensuring that the latest versions of our codes
describe as well as possible fission cross sections from various
reactions. Fig. 3 shows examples of fission cross sections
for proton-induced reactions on 
 $^{186}$W, $^{184}$W, $^{183}$W and $^{182}$W. One can see that
the improved CEM03.01 reproduces the recent Uppsala 
measurements \cite{Eismont05} of proton-induced fission cross 
sections.

Our calculations have shown that CEM03.01 results
also agree reasonably well with the
recent Uppsala \cite{Smirnov05,Smirnov04,Prokofievdata}
and Saint-Petersburg \cite{Fomichev04} data and
older measurements at LANL by Parrish Staples {\it et al.} \cite{Staples}
for neutron-induced fission cross sections.
Results by LAQGSM03.01 for these and other fission cross sections
practically coincide with ones by CEM03.01, as is expected from
the fitting process \cite{fitaf}.

The initial version of LAQGSM \cite{LAQGSM},
just like its precursor QGSM \cite{QGSM}, did not consider photonuclear
reactions, while its 2003 version LAQGSM03 \cite{ND2004} and
CEM03.01 describe such reactions only for energies up to about 1.5 GeV.
This is not convenient when using our codes to solve  problems
for PRAD, NASA, and other high-energy applications 
or to analyze future high-energy measurements at the
Thomas Jefferson National Accelerator Facility (CEBAF),
where photons with much higher energy
are created and need to be simulated by an event generator
in a transport code. To address this problem,
we have extended LAQGSM03 to describe photonuclear reactions
at energies up to tens of GeV. For this, we took advantage of
the high-energy event generators for $\gamma p$ and  $\gamma n$
elementary interactions from the Moscow high-energy
photonuclear reaction model \cite{Iljinov97} kindly sent us
by one of its co-authors, Dr.\ Igor Pshenichnov, as
mentioned previously.
We have incorporated into LAQGSM03 56 channels to consider  
$\gamma p$ elementary interactions during the cascade stage of
reactions, and 56 channels for  $\gamma n$ interactions.
These reaction channels new to LAQGSM03 are listed in 
Table 1 of Ref.\ \cite{ResNote05} together with all
corresponding details. The improved version of our code extended 
in this manner is referred to as LAQGSM03.01; it allows us to describe
photonuclear reactions at energies both below $\sim 1.5$ GeV
where the earlier LAQGSM03 and CEM03.01 work 
(see, {\it e.g.} Fig.\ 11 in \cite{ResNote05}), and at higher photon energies.
Figs.\ 4 and 5 show two examples of results by LAQGSM03.01 for
high-energy photonuclear reactions; more such results may be
found in \cite{ResNote05}. We note that to the best of our 
knowledge, we are able to describe with LAQGSM03.01
the data shown in Figs.\ 4 and 5 (and other similar
reactions analyzed in \cite{ResNote05})
for the first time; we do not know of
any publication or oral presentation where these measurements
were reproduced by a theoretical model, event generator,
or transport code.

We have benchmarked the new versions of our codes, CEM03.01 and 
LAQGSM03.01, on a variety of particle-particle, particle-nucleus, and
nucleus-nucleus reactions at energies from
10 MeV to 800 GeV per nucleon and find that they describe
reactions generally much better than their predecessors.
Two examples of this work are shown in Figs. 6 and 7.
We note that the 400 GeV 
experimental data 
compared in Fig.\ 7 with our LAQGSM03.01 results 
and other similar measurements of 
$K^+$, $K^-$, $\bar p$, d, t, $^3$He, and $^4$He spectra from $^{181}$Ta,
as well as all measured spectra from Cu, Al, and C 
\cite{p400GeV_p}--{\cite{p400GeV_pi}
are described here simultaneously within a single approach
for the first time: Though these data were measured at FNAL 25 
years ago with a hope of revealing some
``exotic" or unknown mechanisms of nuclear reactions leading to
the production of the measured so called ``cumulative"
({\it i.e.}, kinematically forbidden for quasi-free intranuclear 
projectile-nucleon collisions) particles, we do not know
any publication or oral presentation where these measurements
were reproduced simultaneously within a single approach
by a theoretical model, event generator, or transport code.
It is noteworthy that LAQGSM03.01 describes quite well all 
cumulative particle spectra measured in
\cite{p400GeV_p}--{\cite{p400GeV_pi}
in single approach, without any fitting or free parameters, 
and without involving any ``exotic" reaction mechanisms.
This is also true for the photonuclear cumulative proton yields
shown in Fig. 4.
These results do not imply, of course, that the proton-
or $\gamma$-nucleus interaction physics
is completely described by the reaction mechanisms considered
by LAQGSM03.01. Our present results do not exclude some
contribution

\begin{figure}[h]
\begin{minipage}{20pc}
\includegraphics[width=20pc]{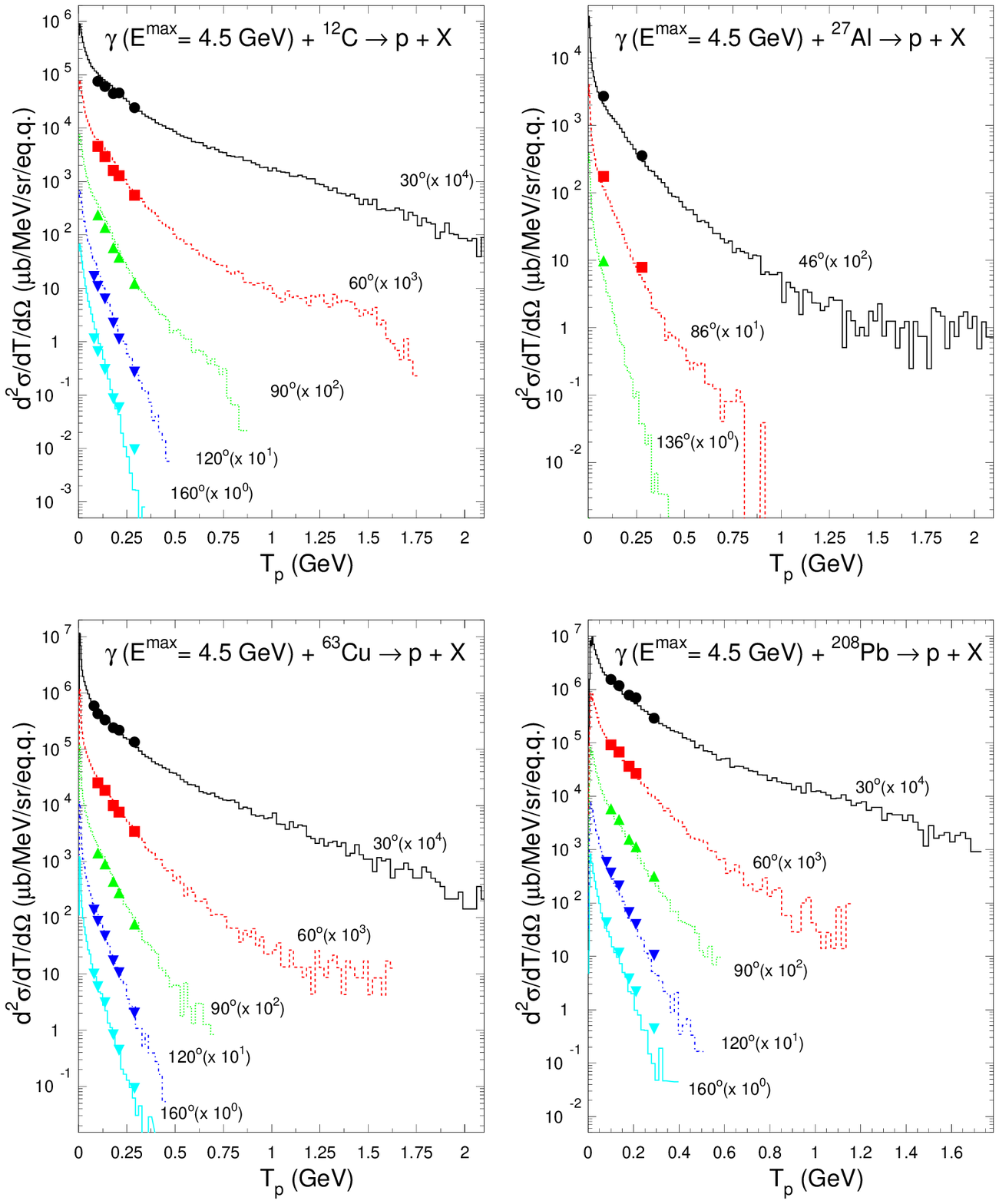}
\caption{\label{label}
Proton spectra at 30, 60, 90, and 150 degrees from
interaction of bremsstrahlung $\gamma$ quanta of maximum energy 
$E_0 = 4.5$ GeV with $^{12}$C,  $^{27}$Al,  $^{63}$Cu, and $^{208}$Pb.
Experimental values shown by symbols are from 
\cite{Alanakyan77,Alanakyan81,Alanakyan81b}
while histograms show results by LAQGSM03.01.
Similar results \cite{ResNote05}
were obtained by  LAQGSM03.01 at $E_0 = 3.0$ and 2.0 GeV,
as well as for photopion spectra
\cite{Egiyan81,Alanakyan81c}. 
To the best of our knowledge, using LAQGSM03.01 we are able to 
describe these data measured 28 years ago at Yerevan for the first time.
}
\end{minipage}
\hspace{1.5pc}
\begin{minipage}{16.4pc}
\includegraphics[width=16.4pc]{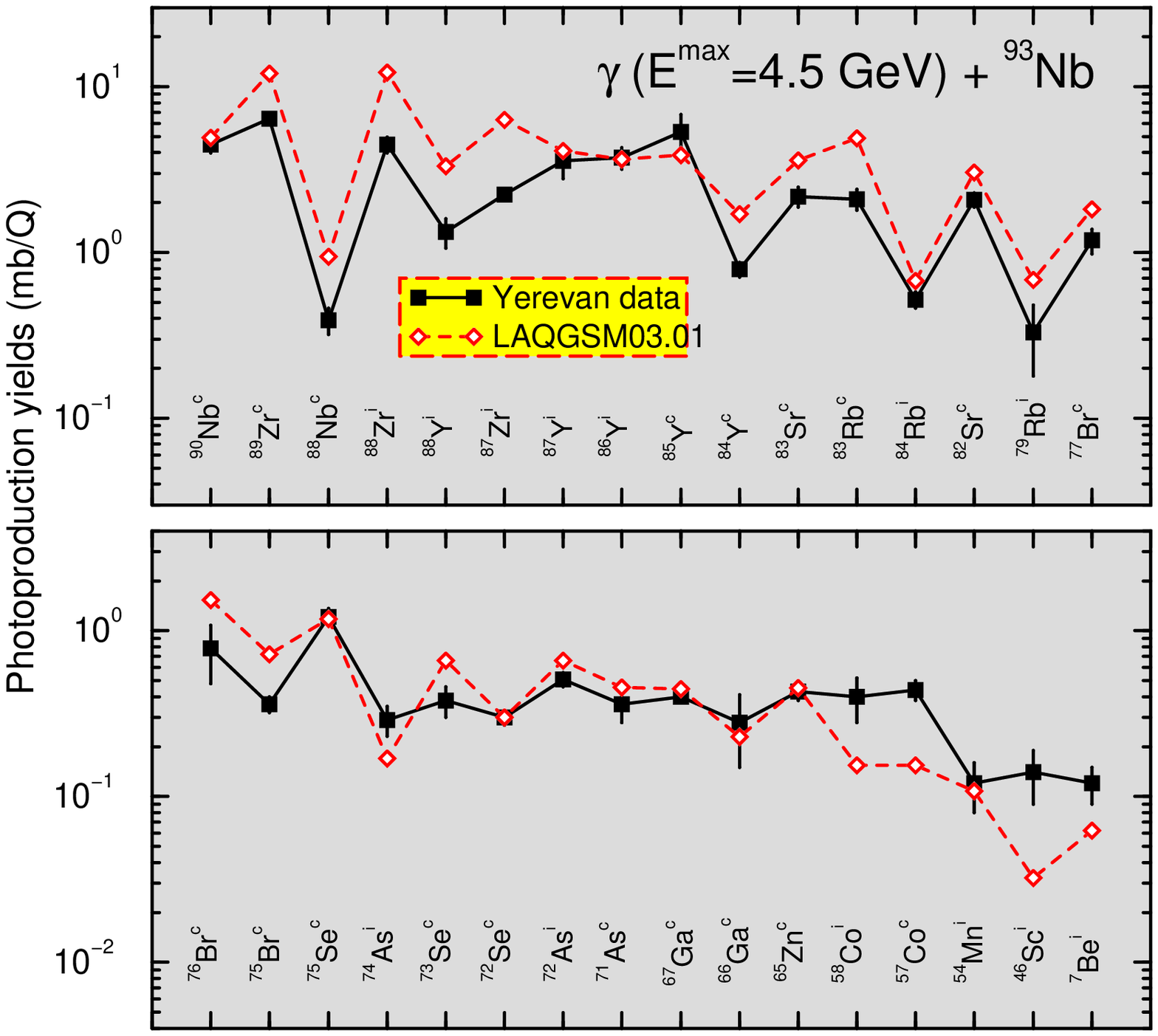}
\caption{\label{label}
Detailed comparison between experimental yields 
\cite{Vartapetyan81}
and those calculated by 
LAQGSM03.01 of radioactive products from the interaction
of bremsstrahlung $\gamma$ quanta of maximum energy 4.5 GeV with 
$^{93}$Nb. The cumulative yields are labeled as ``c" while
the independent cross sections, as ``i".
To the best of our knowledge, using LAQGSM03.01, we are able to 
describe these data measured 24 years ago at Yerevan for the first time.
}
\vspace*{5mm}
\noindent
to the
production of these cumulative
particles from other reaction mechanisms not considered by LAQGSM03.01.
But the contribution from ``exotic" mechanisms to cumulative particle
production from these high-energy reactions seems to be small; inclusive 
particle spectra are not sensitive
\end{minipage} 
\end{figure}

{\noindent
enough for an
unambiguous determination of the mechanisms of particle production,
just as observed heretofore at intermediate and low energies \cite{NP94}.
}

Finally, we wish to note that CEM03.01 and LAQGSM03.01
and their predecessors have good predictive powers
and describe various nuclear reactions similarly to and
often better than other current models do; therefore 
they can be used as reliable event generators 
in transport codes for applications. 
Fig.\ 8 shows just one example to demonstrate this. This figure
was made by adding the recent fission-fragment yield
data from the reactions 1 GeV/A $^{238}$U + d 
 \cite{Priera} (brown circles) to 
Fig.\ A2.3 of Ref.\ \cite{TRAMU}, which showed our
calculations of several models. The calculations were done 
about two years before the data \cite{Priera} became available 
to us (the dates of all our
calculations are shown in the legend of Fig.\ 8).
The prediction by our LAQGSM+GEM2, a predecessor of 
LAQGSM03.01 for the fission-fragment mass distribution from this reaction
agrees very well with the later measurements \cite{Priera}.
The agreement of LAQGSM+GEM2 results
with the older data \cite{Casarejos} (magenta circles) on
spallation product yields (which were available to us before the

\begin{figure}[h]
\begin{minipage}{19.0pc}

\includegraphics[width=19.0pc]{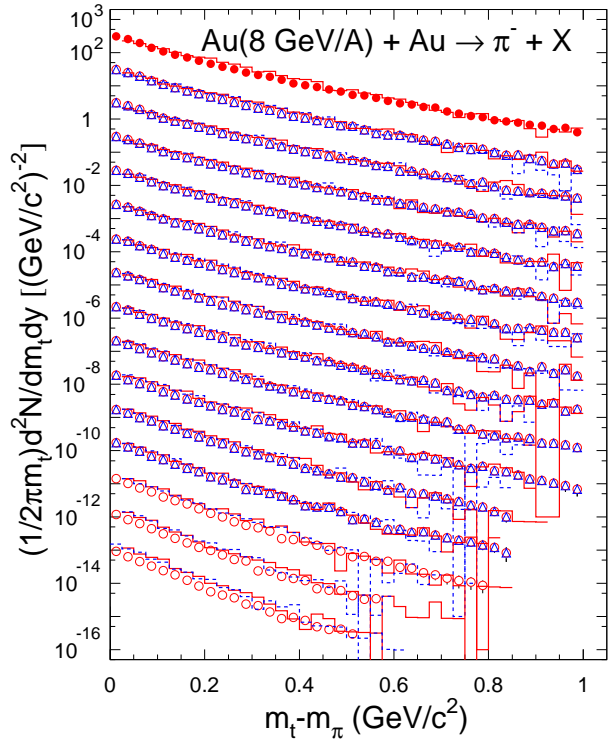}
\caption{\label{label}
Invariant $\pi^-$ yield per central $Au + Au$
collision at 8 GeV/A as calculated by LAQGSM03.01 (histograms)
and measured in \cite{Klay03} (symbols).
Red solid histograms and open circles
are forward production, while blue dashed histograms and open triangles
are backward production. Midrapidity (the upper set) data is shown
unscaled, while 0.1-unit rapidity slices are scaled down by
successive factors of 10.}

\end{minipage}
\hspace{1.0pc}%
\begin{minipage}[h]{17.7pc}

calculations were performed) are also satisfactory
and are comparable to that 
provided by LAHET3 \cite{LAHET3} calculations using: \\
1) ISABEL \cite{ISABEL} INC with the Dresner \cite{Dresner}
evaporation and RAL \cite{RAL} fission models;\\
2) the Liege INC code INCL by Cugnon {\it et al.} \cite{INCL}
merged with the Dresner \cite{Dresner}
evaporation and RAL \cite{RAL} fission models;\\
3) INCL \cite{INCL} merged with 
ABLA/PROFI fission/evaporation model by Schmidt {\it et al.} \cite{ABLA}.

\section{Applications}

CEM03.01 and LAQGSM03.01 and their predecessors are used
as stand-alone codes to study different nuclear reactions for applications 
and fundamental nuclear physics (see, {\it e.g.}, 
\cite{CEM03.01,astro,TRAMU,NP94,MedicalNIM} and references therein). 
We outline below only two examples of applications.

\hspace*{2mm}
The first application of CEM97 and CEM95, predecessors of CEM03.01, we
like to mention here is for a medical isotope production study.
Several years ago, a detailed study of the production of 22 isotopes
for medical and industrial applications by high-energy
protons and neutrons at the Accelerator Production of Tritium 
Facility project (APT) was
performed (see \cite{MedicalNIM,BigMedical} and references therein).
The production rate of a radioisotope can be obtained from the integral of
the flux and cross section leading to the direct production of the
radioisotope as

\end{minipage}
\end{figure}

{\noindent
a reaction product.}
Additional production is realized from
other radionuclides that decay to the desired product. 
Evaluation of
production rates requires knowledge of the neutron and proton fluxes 
at some position in the production facility and cross sections leading to 
production of the desired radionuclide and its progenitors.
We used MCNPX version 2.1.1 to calculate neutron and proton fluxes
throughout the APT model, while cross sections for reactions most
likely to lead to our desired products were evaluated using all
available to us experimental data and calculations by CEM95 and CEM97
at energies above 100 MeV, and the ``150 MeV" activation library
calculated by M. B. Chadwick with the HMS-ALICE code \cite{HMS-ALICE}
supplemented by the European Activation File EAF-97, Rev.\ 1 with
 some recent improvements by
M.\ Herman (see \cite{Herman} and references therein), at lower energies.
Our 684 page detailed report on this study \cite{BigMedical},
with 37 tables and 264 color figures is available on the
Web
under: 
{\bf http://t2.lanl.gov/publications/publications.html}.

A second application of our codes is for
astrophysics. A successful model for cosmic-ray propagation
in the Galaxy used nowadays to study different challenging
questions of astrophysics is realized in the code
GALPROP by Strong and Moskalenko (see \cite{Strong01} 
and references therein).

\begin{figure}[h]
\begin{minipage}{20pc}
\includegraphics[width=20pc]{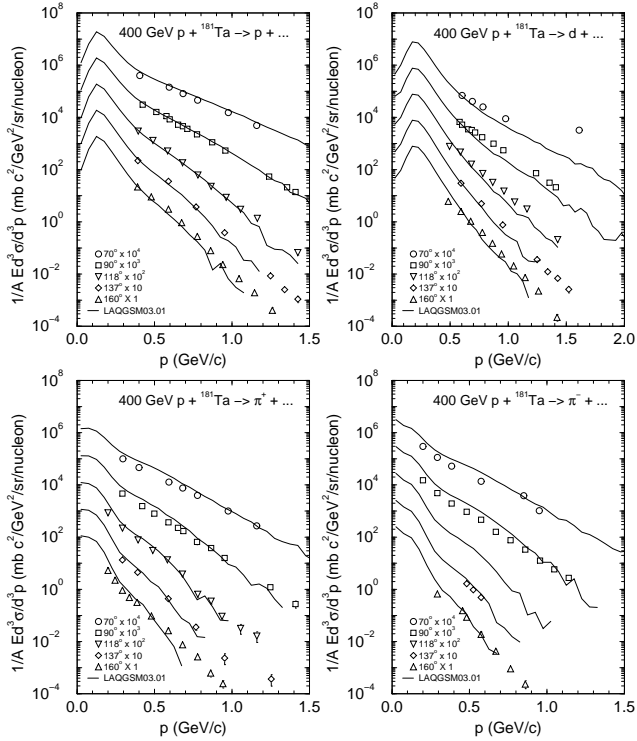}
\caption{\label{label}
Invariant spectra of p, d, $\pi^+$, and $\pi^-$
from the reaction 400 GeV p + $^{181}$Ta.
Experimental data for p are from 
\cite{p400GeV_p}, for d from 
\cite{p400GeV_d}, and for pions from 
\cite{p400GeV_pi}.
Similar results
are obtained by LAQGSM03.01 for the measured 
$K^+$, $K^-$, $\bar p$, d, t, $^3$He, and $^4$He spectra from $^{181}$Ta,
as well as for all measured spectra from Cu, Al, and C 
\cite{p400GeV_p,p400GeV_d,p400GeV_pi}.
To the best of our knowledge, using LAQGSM03.01 we are able to 
describe simultaneously within a single approach
all these data measured 26 years ago at FNAL for the 
first time. It is interesting that we do not need to consider any 
``exotic" nuclear-reaction mechanisms to describe reasonably 
well all these measured ``cumulative" particle spectra.
}
\end{minipage}
\hspace{1.5pc}
\begin{minipage}{16.4pc}

So, GALPROP was used recently
to study propagation of different particles 
and nuclei
in the Galaxy and to
improve estimates of the Galactic halo size
(see, {\it e.g.}, \cite{astro}, \cite{Strong01}-\cite{Ptuskin05}
and references therein).   
However, GALPROP relied initially on nuclear reaction cross sections
calculated using the phenomenological systematics
by Webber {\it et al.} \cite{Weber90}
and Silberberg  {\it et al.} \cite{YIELDX}
and on scarce available experimental data, as 
is usually done in astrophysics.
As an example, Fig.\ 9 shows only one excitation function
for the production of $^{26}$Al from the reaction p + $^{nat}$Si
used by GALPROP  along with hundreds of other excitation functions.
We see that the phenomenological systematics \cite{Weber90,YIELDX}
do not reproduce correctly the available experimental cross
sections for this reaction.
As one can see from Fig.\ 10, using in GALPROP cross sections provided by
phenomenological systematics \cite{Weber90,YIELDX}
leads to quite big uncertainties in the derived Galactic halo size
(3--7 kpc), while using evaluated cross sections based on
available experimental data and calculations by CEM2k \cite{CEM2k}
reduces uncertainties and limits the  derived  Galactic halo size to only
4--6 kpc. This is why we have used several versions of our
CEM and LAQGSM codes to calculate cross sections of interest to astrophysics,
used thereafter together with available experimental data to
produce a number of reliable evaluated excitation functions 
for astrophysical needs
(see, {\it e.g.}, \cite{Moskalenko01}-\cite{Ptuskin05},
\cite{Evaluation} and references therein).     
 
\end{minipage} 
\end{figure}
Finally, we mention that different versions of our CEM and LAQGSM
codes are incorporated wholly as event-generators, or in part in 
different transport codes used in applications,
among them
{\bf CASCADE}~\cite{CASCADE},
{\bf MARS}~\cite{MARS},
{\bf MCNPX}~\cite{MCNPX}
{\bf GEANT4}~\cite{GEANT4},
{\bf SHIELD}~\cite{SHIELD},
{\bf RTS\&T}~\cite{RTSandT},
{\bf SONET}~\cite{SONET},
{\bf CALOR}~\cite{CALOR}, 
{\bf HETC-3STEP}~\cite{HETC-3STEP}, 
{\bf CASCADE/INPE}~\cite{CASCADE/INPE},
{\bf HADRON} \cite{HADRON}, 
{\bf CASCADO}~\cite{CASCADO}, 
{\bf CAMO}~\cite{CAMO}
and others.
Thus our codes are automatically employed in various nuclear applications
where these transport codes are used.

\section{Summary}
Improved versions of the cascade-exciton model of nuclear 
reactions and of the Los Alamos
quark-gluon string model have been developed recently at LANL 
and implemented in the codes CEM03.01 and LAQGSM03.01 
as event-generators for transport codes MCNP6 \cite{MCNP6},
MCNPX \cite{MCNPX}, and MARS \cite{MARS}.
Our codes were previously incorporated into MARS 
and are now being incorporated into MCNPX and  MCNP6.
CEM03.01 was made available to the public via RSICC at Oak Ridge. 
We also plan to make LAQGSM03.01 available to the public via RSICC
in the future.

\ack
This work was supported by the 
Advanced Simulating Computing (ASC) Program at 
the Los Alamos National Laboratory 
operated by the University of California for the
US Department of Energy, and partially by the
Moldovan-US Bilateral Grants Program, CRDF Project MP2-3045,
and the NASA ATP01 Grant NRA-01-01-ATP-066.

\begin{figure}[h]
\includegraphics[width=38pc]{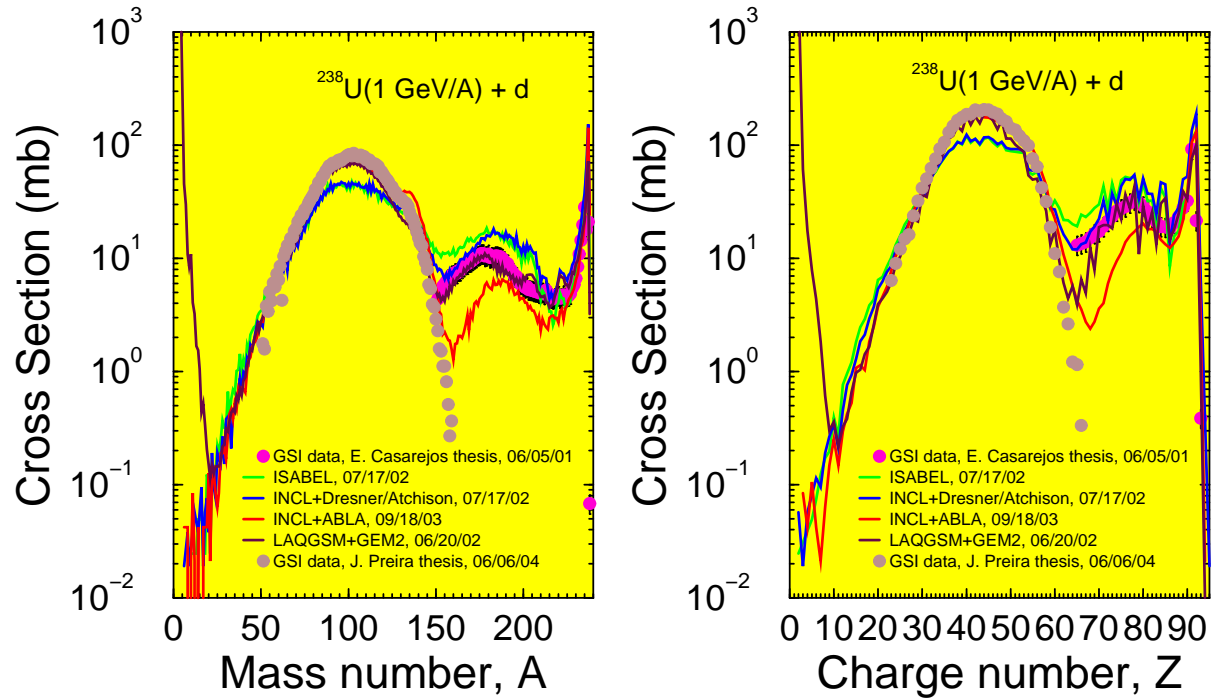}
\caption{\label{label}
Comparison of measured 
mass (left panel) and charge (right panel) distributions
of the nuclides produced in the reactions 
1 GeV/A $^{238}$U + d  with results from LAHET3 \cite{LAHET3}
using the
ISABEL+Dresner/Atchison \cite{ISABEL}-\cite{RAL} (green lines),
INCL+Dresner/Atchison \cite{INCL,Dresner,RAL} (blue line), 
and the INCL+ABLA \cite{INCL,ABLA} (red lines) models, and by our
LAQGSM+GEM2 
(maroon lines), respectively. 
Experimental data for spallation products
shown by magenta circles are
from \cite{Casarejos} and were available to us
prior to the calculations being done at the dates indicated in the
legend, while the fission-product data \cite{Priera}, shown here by
brown circles, became available to us only one year after all the
calculations were published in Fig.\ A2.3 of Ref.\ \cite{TRAMU}.
}
\end{figure}

\begin{figure}[h]
\begin{minipage}{18pc}
\includegraphics[width=18pc]{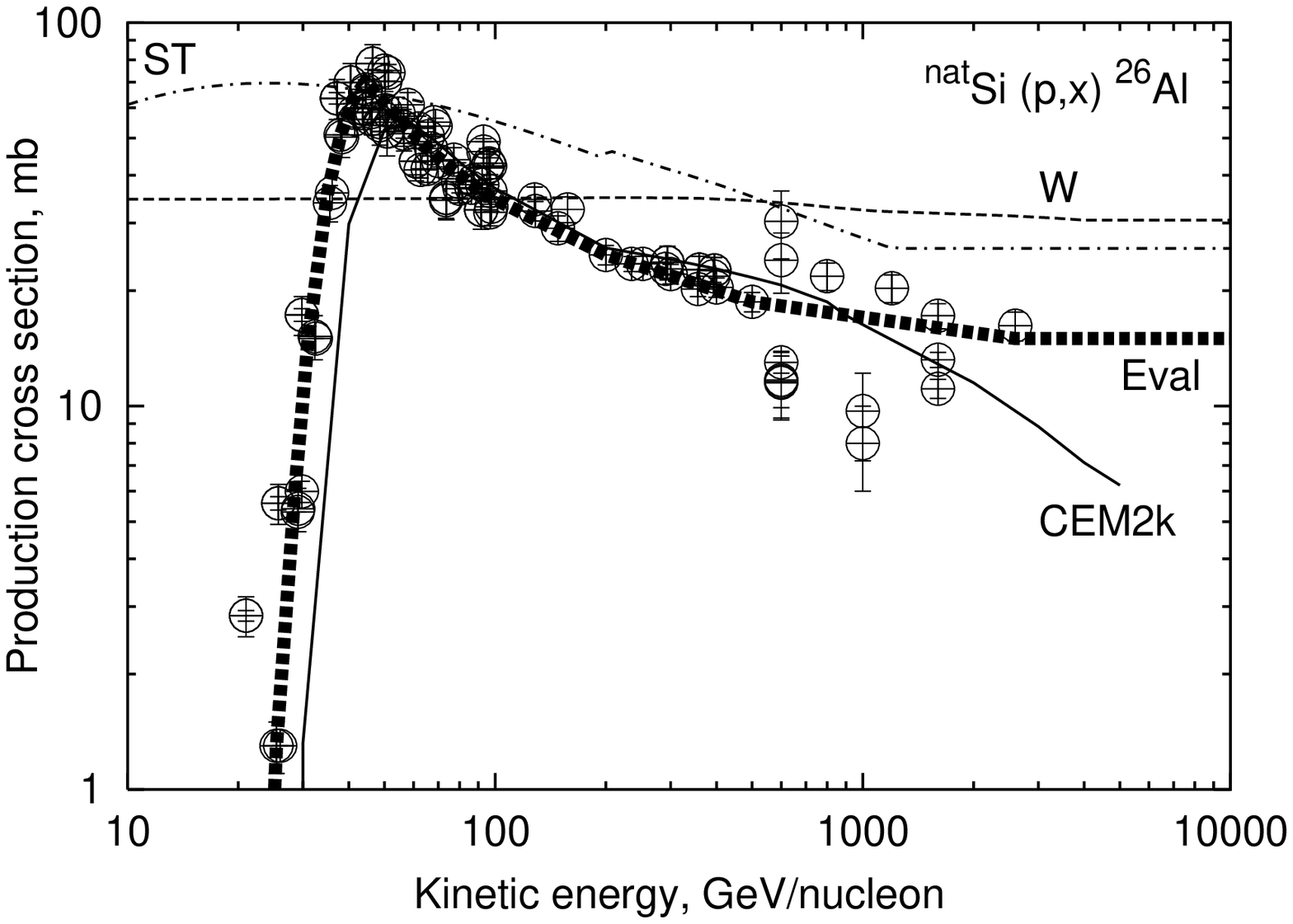}
\caption{\label{label}
Evaluated excitation function for the reaction $^{nat}$Si(p,x)$^{26}$Al
(thick dashed line) compared with experimental data from the LANL T-16 
compilation 
\cite{SARE4c} (our LANL T-16  compilation is 
updated as new experimental data become available to us)
and results from CEM2k 
\cite{CEM2k}
(thin solid line) and phenomenological approximations by Webber {\it et al.} 
\cite{Weber90}
(dashed line) and by Silberberg {\it et al.} 
\cite{YIELDX}
(dot-dashed line).
}
\end{minipage}\hspace{2pc}%
\begin{minipage}{18pc} 
\hspace{1pc}
\includegraphics[width=13pc]{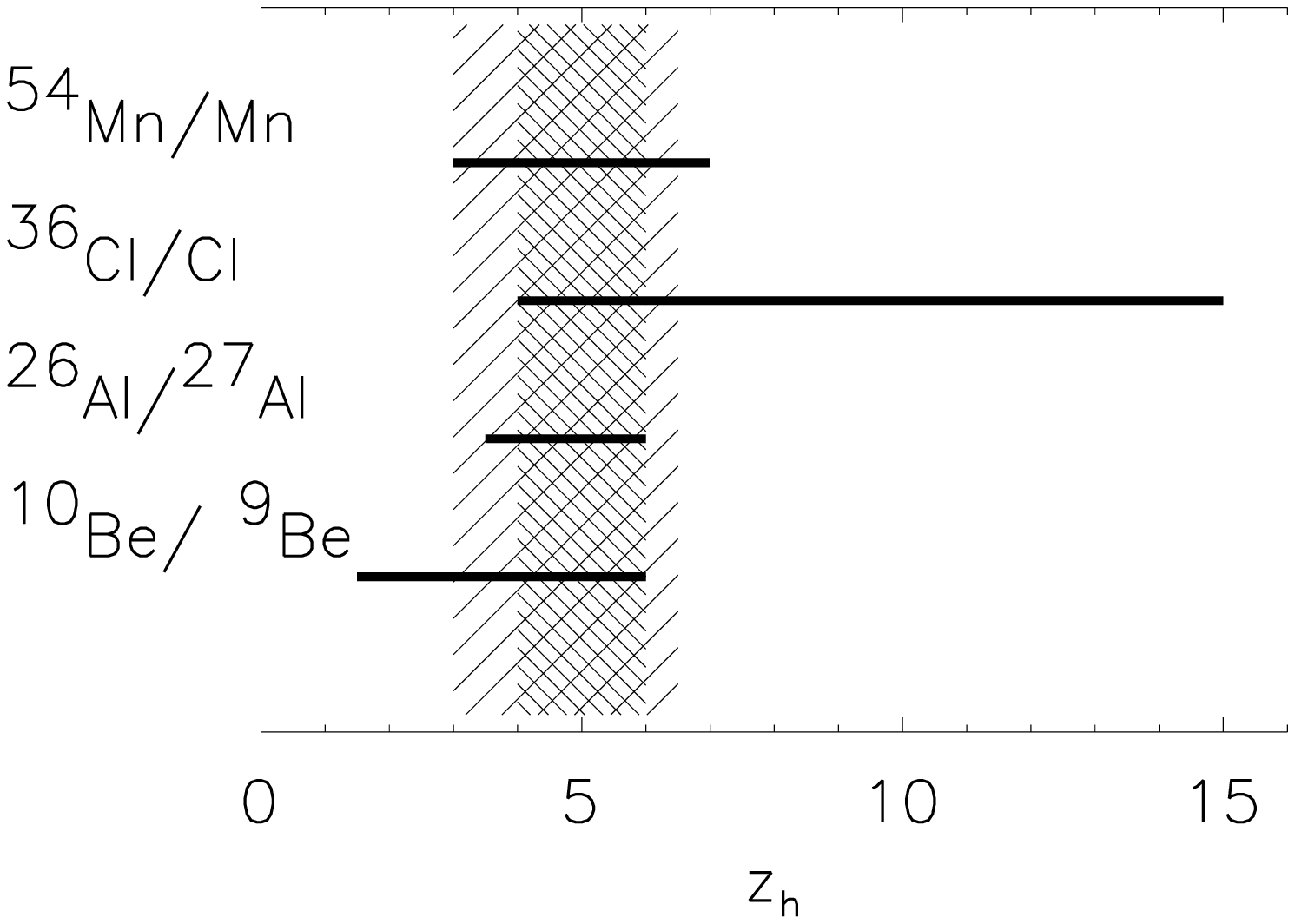}
\caption{\label{label}
Galactic halo size limits  ($Z_h$, kpc) as derived in
\cite{Moskalenko01} from the calculated CR abundances of the four 
radioactive isotopes and ACE spacecraft data.
The ranges given for each isotope were obtained using evaluated
production cross-sections and
reflect errors in measurements of CR isotopic ratios
and CR source abundances. 
The dark shaded area indicates the range consistent with all ratios
(4--6 kpc);
for comparison the range (3--7 kpc) derived 
in \cite{Strong01} using phenomenological approximations 
\cite{Weber90,YIELDX} for cross sections
is shown by light shading.
}
\end{minipage} 
\end{figure}

\end{document}